\documentclass[prb,floatfix,showpacs,amssymb]{revtex4}

\usepackage{graphicx}
\usepackage{dcolumn}
\usepackage{bm}
\begin{document}

\title{Non local Andreev reflection in a carbon nanotube
superconducting quantum interference device}
\author{S. Duhot}
\affiliation{
Institut NEEL, CNRS \& Universit\'e Joseph Fourier, BP 166,
F-38042 Grenoble Cedex 9, France
}
\author{R. M\'elin}
\affiliation{
Institut NEEL, CNRS \& Universit\'e Joseph Fourier, BP 166,
F-38042 Grenoble Cedex 9, France
}

\begin{abstract}
We investigate a superconducting
quantum interference device (SQUID) based on carbon nanotubes
in a fork geometry
[J.-P. Cleuziou {\it et al.}, Nature Nanotechnology {\bf 1}, 53 (2006)], 
involving tunneling of evanescent quasiparticles
through a superconductor over
a distance comparable to the
superconducting coherence length, with therefore ``non local'' processes
generalizing non local Andreev reflection and elastic cotunneling.
Non local processes induce a reduction of the
critical current and modify the current-phase relation. We discuss
arbitrary interface transparencies. Such devices in fork
geometries are candidates
for probing the phase coherence of crossed Andreev reflection.
\end{abstract}

\maketitle
\section{Introduction}\label{sec:Introduction}
Implementing experimentally \cite{Beckmann,Russo,Cadden} 
and understanding theoretically
\cite{Byers,Deutscher,Falci,Samuelson,Prada,Koltai,japs,Feinberg-des,Melin-Feinberg-PRB,Melin-PRB,Levy,Duhot-Melin,Morten,Giazotto,Golubov,Zaikin,Duhot-Melin-cond-mat,Bouchiat,Choi,Martin,Melin-Peysson,Melin-PRB-SQUID} 
the possibility of emitting spatially separated electron pairs
from a superconductor in different electrodes has aroused
considerable interest recently, in connection with the realization
of a source of entangled pairs of electrons \cite{Choi,Martin}.
The Josephson junction through a carbon nanotube quantum dot \cite{Delft}
and the superconducting
quantum interference device \cite{Cleuziou} (SQUID) realized 
recently can be viewed as
steps towards future implementations of
transport of spatially separated pairs of electrons in 
quantum information devices based on carbon nanotubes \cite{Bouchiat}.
As another application, the carbon nanotube SQUID 
realized recently by Cleuziou {\it et al.} \cite{Cleuziou} 
in the fork geometry in Fig.~\ref{fig:squid dc}
proves the
feasibility of future measurements of magnetization reversal of
individual molecular magnets.

Even more miniaturized devices
may be realized in the near future, with
geometrical dimensions comparable to an intrinsic
length scale of the superconductor: the characteristic length\cite{note-xi}
$\xi_0$ associated to the superconducting gap $|\Delta_0|$.
Such devices \cite{Delft,Cleuziou,Cleuziou2}
may be sensitive \cite{Byers,Deutscher,Falci}
to the fact that Andreev reflection
\cite{Andreev}
takes place in a coherence volume of linear dimension $\xi_0$,
therefore allowing for the possibility of splitting
Cooper pairs in two parts of the circuit.

\begin{figure}[htbp]
  \begin{center}
    \includegraphics [width=.3 \linewidth]{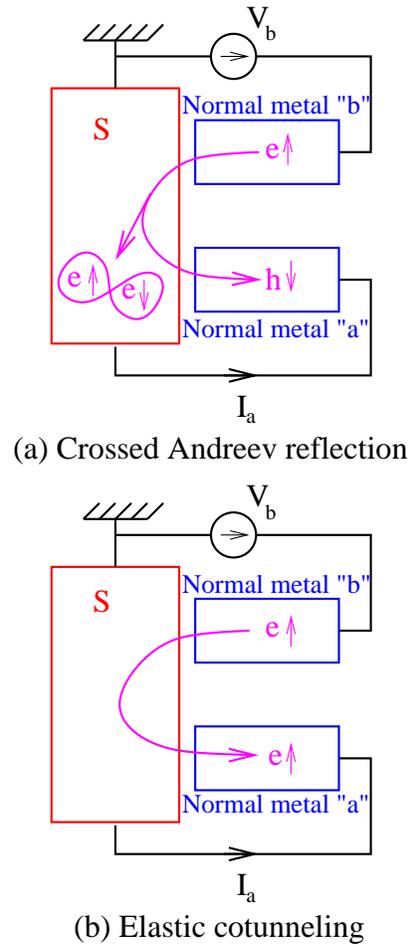}
       \caption{(Color online.) Schematic representation of the
	 electrical circuit for probing the non local conductance in
	 normal metal - superconductor - normal metal 
	 (N$_a$SN$_b$) structures\cite{Delft}.
	 (a) shows a schematic representation of the non local Andreev
	 reflection
	 process changing a spin-up electron in electrode N$_b$ into
	 a spin-down hole in electrode N$_a$ and
	 leaving a Cooper pair in the superconductor. (b) is a schematic
	 representation of elastic cotunneling transferring an electron
	 from one normal electrode to the other via a trip through the
	 superconductor.
}
       \label{fig:CAREC}
  \end{center}
\end{figure}

Andreev reflection \cite{Andreev}
is the process by which a spin-up electron
incoming from the normal side on a NS interface between a normal
metal N and a superconductor S is reflected as a hole in the spin-down
band while a pair of electrons is transmitted in the superconductor. 
In a N$_a$SN$_b$ structure with the electrical circuit on
Fig.~\ref{fig:CAREC},
an electron in electrode N$_a$ can be scattered as
a hole in N$_b$ if the contacts are separated by a distance
comparable to the superconducting coherence length $\xi_0$
(see Fig.~\ref{fig:CAREC}a for non local Andreev reflection).
Alternatively, an electron from N$_a$
can be transmitted as an electron in N$_b$
across the superconductor \cite{Falci}
(see Fig.~\ref{fig:CAREC}b for elastic cotunneling).

The goal of our article is to address possible realizations of
non local Andreev reflection in future carbon nanotube SQUIDs.
These devices\cite{Melin-Peysson,Melin-PRB-SQUID} would provide
further \cite{Delft}
experimental signatures of the phase coherence of
Cooper pair splitting. 
Compared to the previous Refs.~[\onlinecite{Melin-Peysson,Melin-PRB-SQUID}]
we investigate here higher order processes in the tunnel
amplitudes giving rise to ``non local Andreev bound states''. 
By contrast, if the distance between the Josephson junctions is
much larger than the superconducting coherence length, the bound states
are ``local'', and
localized over two separated regions of extend $\xi_0$
on each Josephson junction, not coupled by non local processes
through the superconductor S (see Fig.~\ref{fig:squid dc}).

The article is organized as follows. 
Preliminaries are presented in Sec.~\ref{sec:Preliminaries}.
Our results are presented
in Sec.~\ref{sec:DC} for the dc-Josephson effect 
in single channel systems.
Multichannel effects are discussed in
Sec.~\ref{sec:multi}. 
Concluding remarks are presented in
Sec.~\ref{sec:Conclusions}. Some details are left for Appendices.

\section{Preliminaries}\label{sec:Preliminaries}

\subsection{Carbon nanotube superconducting interference device}
\begin{figure}[htbp]
  \begin{center}
    \includegraphics [width=.8 \linewidth]{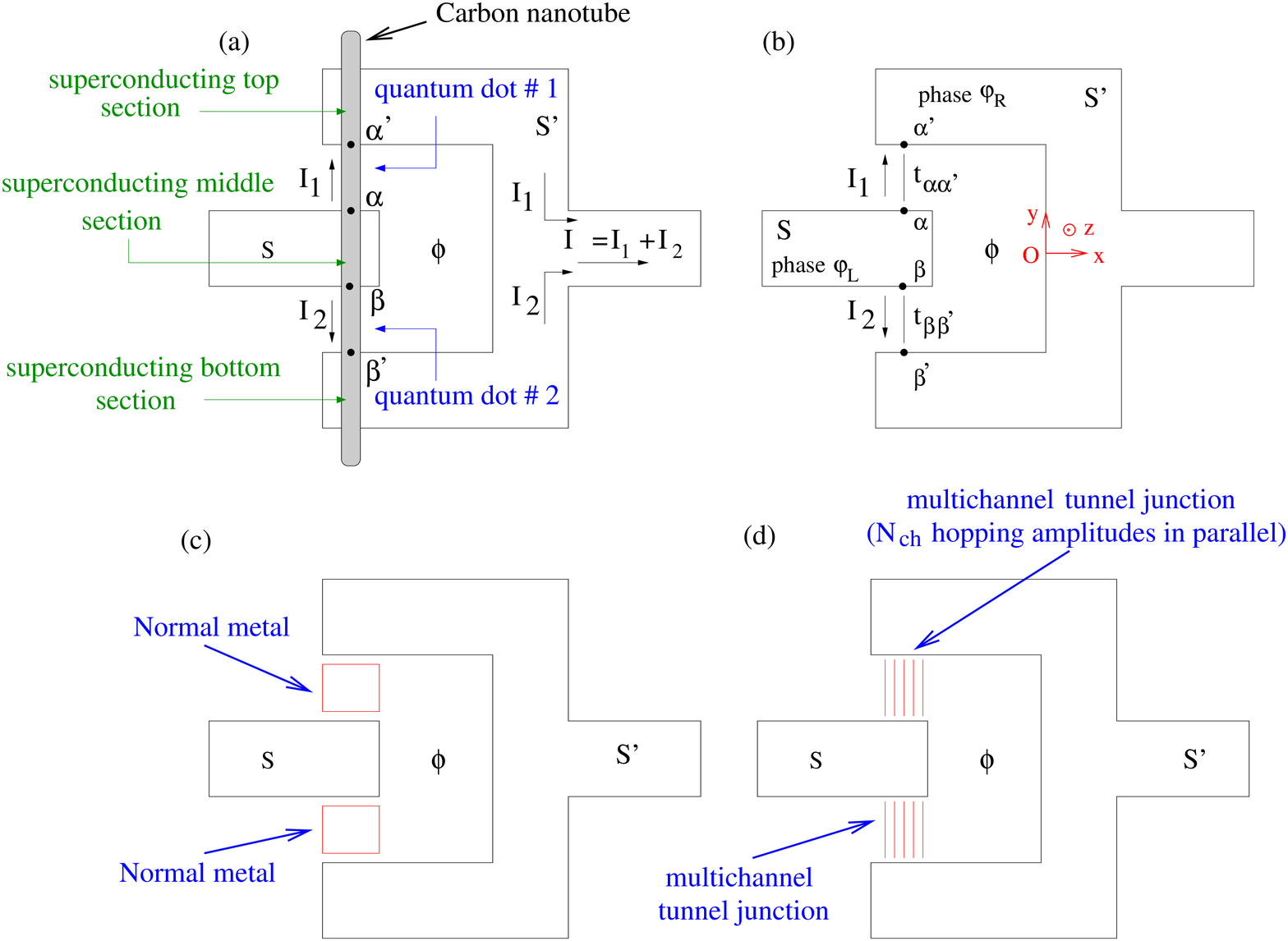}
       \caption{(Color online.) Schematic representation of the
	 carbon nanotube  SQUID (a). The hopping
	 description in the off-resonant state is shown on (b).
	 The transport properties of the SQUID depend only on the
	 enclosed flux and on the superconducting phase differences,
	 (c) shows a double bridge
	 between two superconductors.
	 (d) is a double insulating bridge between two superconductors,
	 made of two multichannel insulators in parallel. Only 
	 non local
	 processes through one of the superconductors (the superconductor S)
	 are allowed in all these situations.
	 The $Oxyz$ axis is shown on (b).
       }
       \label{fig:squid dc}
  \end{center}
\end{figure}

A superconducting quantum interference device (SQUID) made of two
superconductors and a carbon nanotube is considered (see
Fig.~\ref{fig:squid dc}a), according to the recent
experiment by Cleuziou {\it et al.}
\cite{Cleuziou}. As a natural hypothesis, the proximity
effect between the superconductor and the carbon nanotube 
({\it i.e.} the penetration of pairs from the superconductor
to the nanotube) is supposed to induce a minigap $|\Delta_0|$
in the portions of the nanotube in contact with the
superconductors. 

The nanotube is divided in five sections 
connected to each other, from top to bottom:
superconducting top section with a minigap
$|\Delta_0|$ in contact with the superconductor S';
quantum dot number 1; superconducting middle section
of the nanotube with a minigap $|\Delta_0|$ in contact with S;
quantum dot number 2;
and superconducting bottom section with a minigap
$|\Delta_0|$ in contact with the superconductor S'
(see Fig.~\ref{fig:squid dc}a).

\begin{figure*}[htbp]
  \begin{center}
    \includegraphics [width=.7 \linewidth]{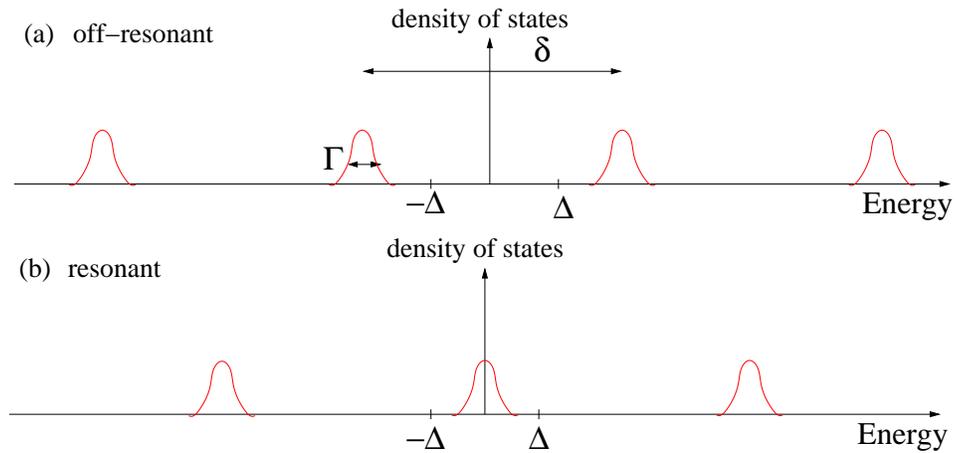}
       \caption{(Color online.) 
	 Schematic representation of quantum dot density of states,
       with a dot level spacing $\delta$ and a level broadening $\Gamma$.
       (a) corresponds to the off-resonant state considered in our article,
       with no resonant level in the gap window. (b) corresponds to the
       resonant situation with a large density of states within the
       gap. 
}
       \label{fig:reso}
  \end{center}
\end{figure*}

Depending on the value of the gate voltage in experiments,
the quantum
dots 1 and 2 can be tuned from
off-resonant to resonant (changing the gate voltages has
the effect of shifting the dot energy levels).
Our article discusses mostly the off-resonant state (in short:
off-state) such that dots 1 and 2 have a vanishingly small
density of states within the minigap (see Fig.~\ref{fig:reso}a).
It was well established experimentally by Cleuziou {\it et al.}\cite{Cleuziou}
that their SQUID can be tuned from the off-state with a very small critical
current to the on-state with a large critical current by changing the
gate voltages coupled to the two quantum dots formed by portions of the
nanotube in between $\alpha$ and
$\alpha'$, and in between $\beta$ and $\beta'$ (see Fig.~\ref{fig:squid dc}a).

Our modeling is intended to capture ``non local bound states'' involving
multiple electron-hole processes back and forth between dots 1 and 2.
We make the following simplifying assumptions.
First, proximity
effect between the nanotube and the superconductor is not
described explicitely on a microscopic basis: for
highly transparent interfaces between the carbon nanotube and
the superconductor S,
we treat proximity-induced superconductivity
in the nanotube as bulk superconductivity, and therefore we use the
denomination ``gap'' instead of ``minigap''.
Non local processes
then take place at the discontinuities
of the superconducting order parameter at $\alpha$ and
$\alpha'$.
For lower interface transparencies between the superconductor S
and the carbon nanotube, electrons and holes may
propagate in the portion of the nanotube in contact with S
before undergoing non local Andreev reflection or elastic
cotunneling, which amounts to
averaging over many channels for non local processes.

Second, we consider the off-state with a vanishingly small
density of states if the absolute value of energy is
smaller than the gap (see Fig.~\ref{fig:reso}a), and we use
a standard description
as tunnel amplitudes \cite{Caroli,Cuevas,Averin} connecting
the superconductors S and S'. 

As a third assumption, it is well known that
a single wall carbon nanotube contains two conduction channels.
The idealized case of a single hopping amplitude is discussed
in Sec.~\ref{sec:DC}.
Multichannel contacts are left for Sec.~\ref{sec:multi}.

As a final assumption, Coulomb interactions in the quantum dots
are not accounted for, so that $\delta$ on Fig.~\ref{fig:reso}
is supposed to be
a finite size effect not due to Coulomb interactions.

The distance $R_{\alpha,\beta}$ between $\alpha$ and $\beta$ (see
$\alpha$ and $\beta$ on
Fig.~\ref{fig:squid dc}b) is supposed to be comparable to the 
coherence length $\xi_0$.
The shortest path connecting $\alpha'$
and $\beta'$ (see $\alpha'$ and $\beta'$ on Fig.~\ref{fig:squid dc}b)
is much larger than $\xi_0$ ($R_{\alpha',\beta'}\gg \xi_0$), with therefore
no ``non local'' quasiparticle tunneling between $\alpha'$
and $\beta'$. Compared to the fork in the experimental
geometry realized in Ref.~[\onlinecite{Cleuziou}],
these assumptions on geometry can be implemented in future
experiments by reducing the distance between $\alpha$ and $\beta$
(see $\alpha$ and $\beta$ on Fig.~\ref{fig:squid dc}b) down to values
comparable to the superconducting coherence length $\xi_0$.
In the experiment by Cleuziou {\it et al.}\cite{Cleuziou},
the distance $R_{\alpha,\beta}$
between $\alpha$ and $\beta$ is comparable to the
distance $R_{\alpha,\alpha'}$
between $\alpha$ and $\alpha'$, and to the distance
$R_{\beta,\beta'}$ between
$\beta$ and $\beta'$ (of order $400$~nm).
We consider on the contrary future fork geometries with 
$R_{\alpha,\beta}\agt\xi_0$, and
with
$R_{\alpha,\beta}\ll R_{\alpha,\alpha'},R_{\beta,\beta'}$.

Choosing the gauge $A_x=-By/2$, $A_y=Bx/2$
and $A_z=0$ (with the $Oxyz$ axis on Fig.~\ref{fig:squid dc}b),
with $B$ the applied magnetic field and $\bf{A}$ the vector
potential, leads to a finite value for
$\int_{\alpha}^{\beta}\bf{A}\mbox{d}\bf{r}$,
and to $\int_{\alpha}^{\alpha'}\bf{A}\mbox{d}\bf{r}=
-\int_{\beta}^{\beta'}\bf{A}\mbox{d}\bf{r}$ if we
suppose $R_{\alpha,\alpha'}=R_{\beta,\beta'}$.
Given $R_{\alpha,\beta}\ll R_{\alpha,\alpha'}$,
we neglect the line integral of the vector potential
between $\alpha$ and $\beta$:
$\int_{\alpha}^{\beta}{\bf A}\mbox{d}{\bf r}\simeq0$.
The line integral of the vector potential along a path from
$\alpha'$ to $\beta'$ in S' is finite but quasiparticle
propagation from $\alpha'$ to $\beta'$ has a vanishingly small
amplitude (because $R_{\alpha',\beta'}\gg \xi_0$).
Non local processes between $\alpha'$ and $\beta'$ are thus negligible.

\subsection{Microscopic Green's functions}

The SQUID corresponds to two Josephson junctions, one for each interface.
If the junctions are far apart (at a distance much larger than the
superconducting coherence length), and for single channel weak links,
one negative and one positive
energy Andreev bound state is located on each junction, therefore
leading to 
a total of four Andreev bound states for the SQUID
(two bound states at positive energy with respect to the Fermi level,
and two bound states at negative energy).
As we show below, non local processes induce a coupling between these
bound states in the form of level repulsion.

The supercurrent is obtained from differentiating the free energy with respect
to the superconducting phase difference $\Delta\varphi$. 
Beenakker\cite{Beenakker} finds three terms
contributing to the supercurrent, some of which date back to the
early stages of Josephson junction theory \cite{Bardeen}.
First at zero temperature
the following term corresponds to a summation over the 
discrete bound states within the gap:
\begin{eqnarray}\label{eq:thermo}
I_S(\Delta\varphi,\phi)=\frac{2e|\Delta_0|}{\hbar}
\sum_{n=1}^{N_{ABS}} \frac{\partial
  \Omega_n(\Delta\varphi,\phi)}{\partial(\Delta\varphi)}
\theta\left[-\Omega_n(\Delta\varphi,\phi)\right]
,
\end{eqnarray}
where $\phi$ is the
flux enclosed in the loop of the SQUID, and where $N_{ABS}$ is the number of
Andreev bound states. The step function in energy
$\theta\left[-\Omega_n(\Delta\varphi,\phi)\right]$ selects Andreev
bound states below the Fermi level.
The second term in Ref.~[\onlinecite{Beenakker}], 
corresponds to
the contribution of the continuum to the supercurrent.
The contribution of the continuum is not included in
the discussion in the forthcoming
Secs.~\ref{sec:DC} and~\ref{sec:multi}.
We will justify in
Sec.~\ref{sec:continuum} that it is indeed negligibly small in the
situations that we consider. The third and last term
in the expression of the supercurrent obtained by Beenakker\cite{Beenakker}
vanishes if the superconducting gap is independent
on the phase difference, which we assume in the following.

The bound states are obtained in a standard description as the poles
of the fully dressed Green's functions. The later is determined by
the Dyson equations, which allows to describe weak links ranging from
tunnel contacts to highly transparent interfaces. 

The Green's functions of the superconductor are
obtained by
Fourier transform in a well known procedure \cite{AGD}.
For superconductors isolated from each other,
the local Green's function takes the form
\begin{eqnarray}
\hat{g}_{\alpha,\alpha}(\omega)&=&\hat{g}_{\beta,\beta}(\omega) =
\frac{\pi\rho_N}{\sqrt{|\Delta_0|^2-(\hbar\omega)^2}}
\left(\begin{array}{cc}
  -\hbar\omega & |\Delta_0|\exp(i\varphi_L) \\
|\Delta_0|\exp(-i\varphi_L) & -\hbar\omega 
\end{array}\right),\\
\hat{g}_{\alpha',\alpha'}(\omega)&=&\hat{g}_{\beta',\beta'}(\omega)
=
\frac{\pi\rho_N}{\sqrt{|\Delta_0|^2-(\hbar\omega)^2}}
\left(\begin{array}{cc}
  -\hbar\omega & |\Delta_0|\exp(i\varphi_R) \\
|\Delta_0|\exp(-i\varphi_R) & -\hbar\omega 
\end{array}\right),
\end{eqnarray}
where the superconducting phase
variables $\varphi_L$ and $\varphi_R=\varphi_L+\Delta\varphi$
are shown on Fig.~\ref{fig:squid dc}b, $|\Delta_0|$ is the
superconducting gap and $\rho_N$ the normal density of states.

The non local Green's functions take the form
\begin{eqnarray}\label{gab 1D}
\nonumber
g_{\alpha,\beta}(\omega)&=&g_{\beta,\alpha}(\omega)
={\cal C}(\omega)\pi\rho_N\left\{\frac{1}{\sqrt{|\Delta_0|^2-(\hbar\omega)^2}}
\left(\begin{array}{cc}
  -\hbar\omega & |\Delta_0|\exp(i\varphi_L) \\
|\Delta_0|\exp(-i\varphi_L) & -\hbar\omega 
\end{array}\right)
\cos{(k_F R_{\alpha,\beta})}\right.\\ 
&&\left.+\left(\begin{array}{cc}
  -1 & 0 \\
0 & 1 
\end{array}\right)\sin{(k_F R_{\alpha,\beta})}
\right\}
,
\end{eqnarray}
where we use the notation ${\cal C}(\omega)$ for
\begin{equation}
\label{eq:C-omega}
{\cal C}(\omega)=\exp{\left[-2 \frac{R_{\alpha,\beta}}{\xi(\omega)}
\right]}
.
\end{equation}
We 
parameterize below the strength of non local processes by
${\cal C}_0={\cal C}(\omega=0)=\exp{[-2 R_{\alpha,\beta}/\xi_0]}$,
with $\xi_0=\xi(\omega=0)$.
The strength of non local processes is parameterized by ${\cal C}_0$,
ranging from an absence of non local processes (${\cal C}_0\simeq 0$
for $R_{\alpha,\beta}\gg\xi_0$) to 
non local processes taking their maximal value (${\cal C}_0\simeq 1$
for $R_{\alpha,\beta}\ll\xi_0$). One has then the following:
\begin{equation}
{\cal C}(\omega)=\left[{\cal C}_0\right]^{\xi(\omega)/\xi_0}
.
\end{equation}
Intermediate values of 
${\cal C}_0$ are
expected for a carbon
nanotube with $R_{\alpha,\beta}$ of order $\xi_0$.
Note that in the case of
three dimensions (not considered here), the $\cos{(k_F R_{\alpha,\beta})}$ and
$\sin{(k_F R_{\alpha,\beta})}$ factors are interchanged,
and
${\cal C}(\omega)=
\exp{[-2 R_{\alpha,\beta}/\xi(\omega)]}/R_{\alpha,\beta}$.

The condition $R_{\alpha',\beta'}\gg \xi_0$ (see Fig.~\ref{fig:squid dc}b)
leads to
\begin{eqnarray}
g_{\alpha',\beta'}(\omega)=g_{\beta',\alpha'}(\omega)=\left(\begin{array}{cc}
  0 & 0 \\
0 & 0 
\end{array}\right)
.
\end{eqnarray}

The fully dressed Green's functions at energy $\omega$
are obtained via the Dyson equations taking the following form in a compact
notation:
\begin{equation}
\hat{G}(\omega)=\hat{g}(\omega)+\hat{g}(\omega)\otimes\hat{\Sigma}_t
\otimes\hat{G}(\omega)
,
\end{equation}
where $\hat{g}(\omega)$ corresponds to the Green's functions of the
superconducting electrodes isolated from each other in the absence 
of tunnel amplitudes, $\hat{\Sigma}_t$ is the Nambu hopping self-energy,
and $\hat{G}(\omega)$ is the fully dressed Green's function.
The notation $\otimes$ denotes a convolution over the network labels
$\alpha$, $\beta$, $\alpha'$ and $\beta'$
(see the notations on Fig.~\ref{fig:squid dc}b). For instance one has
the following:
\begin{equation}
\hat{G}_{\alpha,\beta}(\omega)=\hat{g}_{\alpha,\beta}(\omega)
+\hat{g}_{\alpha,\alpha}(\omega) \hat{t}_{\alpha,\alpha'}
\hat{G}_{\alpha',\beta}(\omega)+
\hat{g}_{\alpha,\beta}(\omega)\hat{t}_{\beta,\beta'}
\hat{G}_{\beta',\beta}(\omega).
\end{equation}
The set of fully dressed Green's functions are then obtained 
from matrix inversion, and the Andreev bound states correspond to the
poles within the gap. They are determined either from the corresponding
analytical expressions of the Green's functions, or from a numerical
solution.

\label{sec:toto1}

\section{Results}
\subsection{DC Josephson effect for a
single transmission channel}\label{sec:DC}

\subsubsection{Amplitude and minima 
of the critical current}\label{subsec:Amplitude}

\begin{figure}[htbp]
  \begin{center}
    \includegraphics [width=.6 \linewidth]{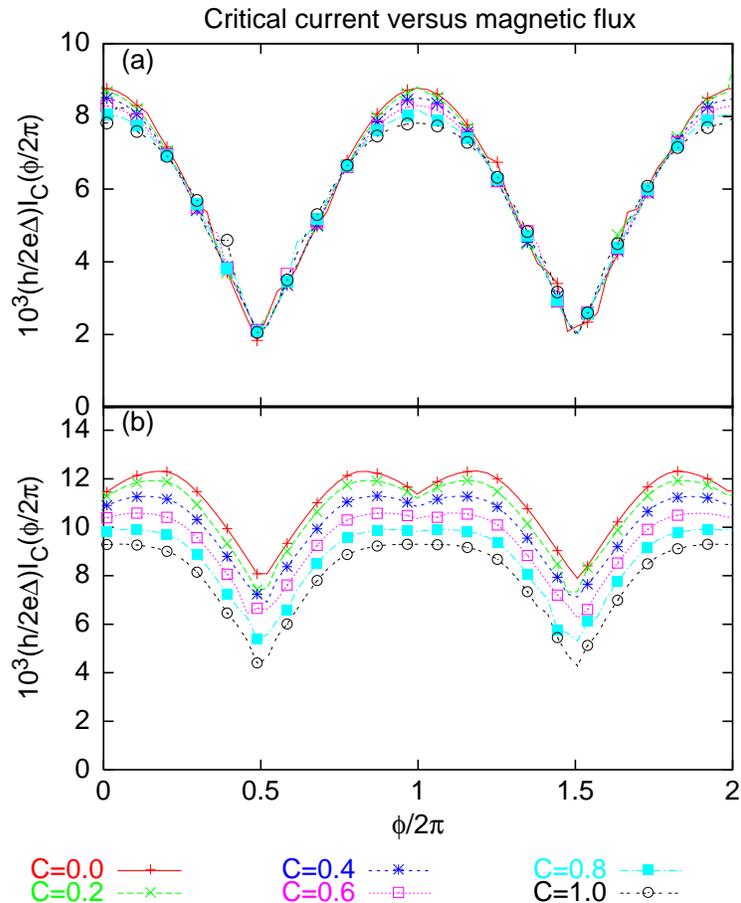}
       \caption{(Color online.) 
	 Critical current as a function of the
	 magnetic flux $\phi$ in the loop of the SQUID, with
	 $(T_\alpha,T_\beta)=(0.8,0.8)$ (a) and 
	 with $(T_\alpha,T_\beta)=(0.8,1)$ (b). The critical current is
	 averaged over all values of $k_F R_{\alpha,\beta}$.
	 Different curves correspond to different values of ${\cal C}_0$,
	 the strength of non local processes. 
}
       \label{fig:Critical}
  \end{center}
\end{figure}

We find a reduction of the
supercurrent upon increasing the strength
${\cal C}_0$
of non local processes (see Fig.~\ref{fig:Critical}). On
this figure, the critical current is averaged over all
realizations of the Fermi phase factor $k_F R_{\alpha,\beta}$
corresponding to averaging over many samples with
different Fermi phase factors.
The interfaces of a superconducting electrode 
are not controlled on atomic scale
and it is thus a natural assumption \cite{Falci,Hekking}
to use a uniform distribution of the Fermi phase factors
$k_F R_{\alpha,\beta}$. 
As expected,
the reduction of the supercurrent by non local processes 
in a collection of single channel systems
is
in agreement with a collection of multichannel systems
(see Sec.~\ref{sec:multi}).

Anticipating the forthcoming Sec.~\ref{sec:expli},
we note that non local Andreev reflection changes an
electron with positive energy on one junction into a hole with negative
energy on the other junction. As a consequence, bound states with positive
energy are coupled to bound states with negative energies. 
The resulting level repulsion
among bound states with opposite energies 
reduces in absolute value the slope of the phase dependence of the Andreev
levels and therefore reduces the critical current.
We evaluate in Appendix~\ref{app:pert} the critical current for
tunnel interfaces, and we confirm by this analytical treatment the
reduction of the critical current by non local processes.

\subsubsection{Level repulsion among Andreev bound states and 
current-phase relation}\label{subsec:Bound}
\label{sec:expli}

Assuming a distance between the Josephson
junctions much larger than the coherence length, and assuming also single
channel contacts, we find
one Andreev bound state with negative energy localized
on each junction, as it should.
The bound states extend in the superconductor over a
region of size comparable to the coherence length. If the
distance between the Josephson junctions 
becomes comparable to the coherence
length, 
the bound state energy levels depend on the coupling corresponding to "non
local" propagation in the superconductors (see Fig.~\ref{fig:squid dc}).

\begin{figure}[htbp]
  \begin{center}
    \includegraphics [width=.45 \linewidth]{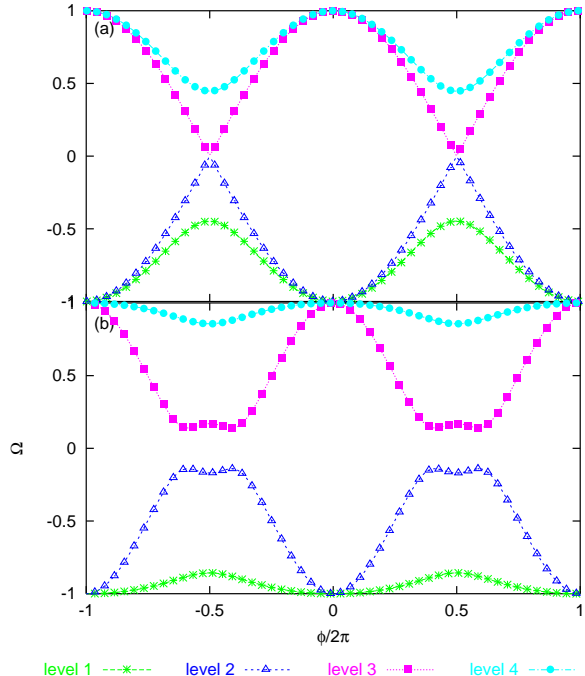}
       \caption{(Color online.) Variation of the bound state
       levels in the absence of non local processes 
       (${\cal C}_0=0$)(a) and with the maximal strength of non local
       processes
       (${\cal C}_0=1$) (b). The bound states repel each other 
       for ${\cal C}_0=1$ in the
       presence of non local processes on (b). For this figure,
       we use $\Delta\varphi=0$,
       $k_F R_{\alpha,\beta}=1+2\pi n$ (with $n$ an integer),
       $T_\alpha=0.8$ and $T_\beta=1$. A similar repulsion between
       Andreev bound states is obtained in the dependence of the
       bound state energy levels as a function of the phase difference
       $\Delta\varphi$.
}
       \label{fig:Energy}
  \end{center}
\end{figure}

The variations of the bound state levels ($\pm \Omega_1$, $\pm
\Omega_2$) with the flux $\phi$ enclosed in the loop
are shown on Fig.~\ref{fig:Energy}a for
${\cal C}_0=0$ (absence of non local processes) and 
on Fig.~\ref{fig:Energy}b for ${\cal C}_0=1$
(maximal value of non local processes). 

Level repulsion among Andreev
bound states upon increasing ${\cal C}_0$ (see Fig.~\ref{fig:Energy}b)
has the effect of reducing the slope of the bound state energy levels
versus phase relation, which reduces the supercurrent.
The bound
states take the simple form given in
Appendix~\ref{sec:Appendice B} (see Eqs. (\ref{omega sym pm}) and 
(\ref{omega antisym pm}))
for a symmetric contact with $t_a=t_b$. We do not present in
the article the too heavy expression of the bound state levels in
the general case.

\label{subsec:Current-phase}

\begin{figure}[htbp]
  \begin{center}
    \includegraphics [width=.5 \linewidth]{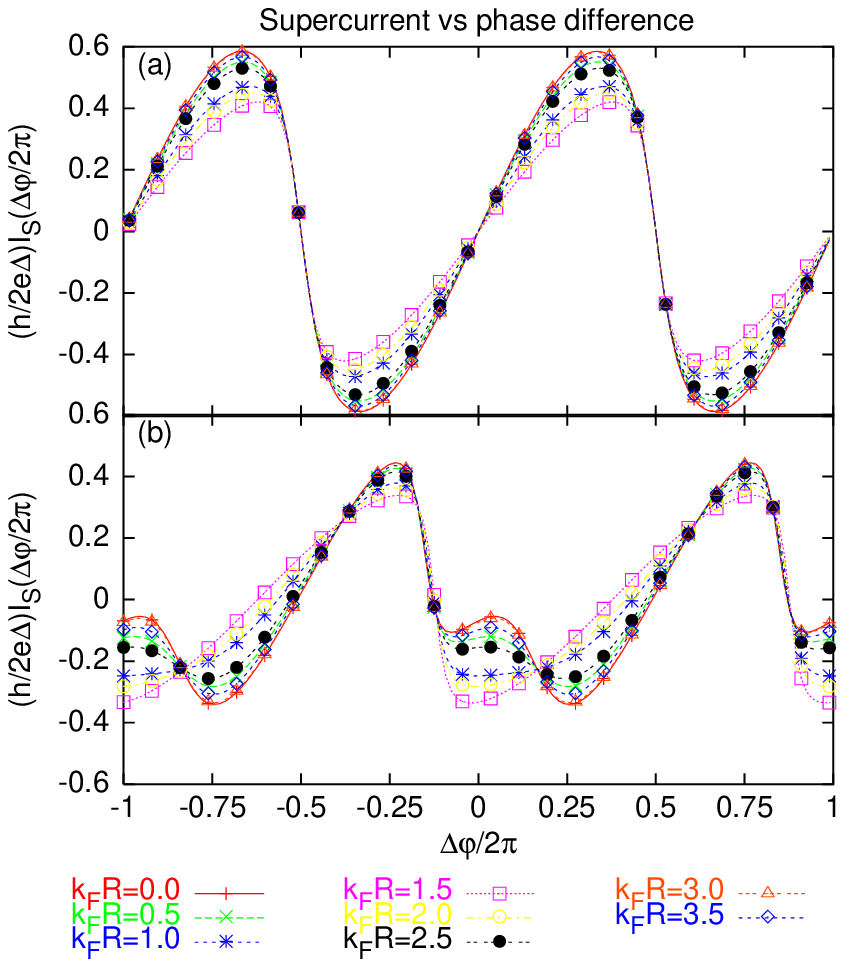}
    \caption{(Color online.) 
      Supercurrent versus phase
      difference $\Delta\varphi$, with non local processes, for
      different values of $k_F R_{\alpha,\beta}$, and for
      $(T_\alpha,T_\beta)=(0.8,0.8)$, 
      without magnetic flux $\phi=0$ (a), with magnetic flux
      $\phi=2$ (b). The supercurrent is very small for
       $k_F R_{\alpha,\beta}=1.5+2\pi n$ (with $n$ and integer),
      not far from $k_F R_{\alpha,\beta}\simeq \pi/2+2\pi n$.}
\label{fig:graph}
  \end{center}
\end{figure}

Now we consider 
single channel transmission modes between the superconductors
S and S', and with non local processes at the interfaces
where a step function variation of the superconducting gap is
assumed.
As seen from Fig.~\ref{fig:graph}, the
SQUID current phase relation fluctuates from sample to sample.

\label{sec:continuum}

\begin{figure}[htbp]
  \begin{center}
    \includegraphics [width=.6 \linewidth]{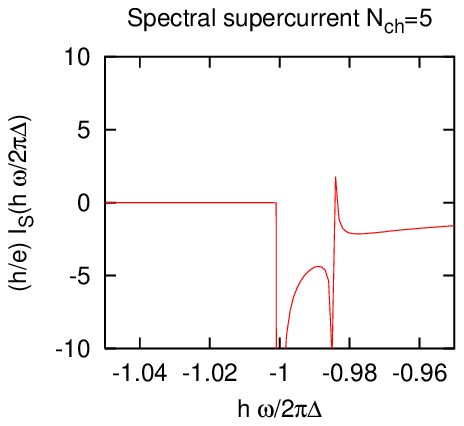}
       \caption{(Color online.)
	 Variations of the spectral supercurrent for a given
	 realization of the microscopic Fermi phase factors in the
	 vicinity of $\hbar \omega=h\omega/2\pi
	 \simeq-|\Delta_0|$. We find practically
	 no contribution of the continuum for the hopping model
	 of SQUID in the
	 off-state. We use the phase difference $\Delta\varphi=2$
	 on the figure, but similar variations of the spectral
	 current are obtained for other values of $\Delta\varphi$.
	 We also chose ${\cal C}_0=1$ on the figure. 
       }
       \label{fig:continuum}
  \end{center}
\end{figure}
We conclude this section by discussing
the contribution of the continuum \cite{Beenakker} for
the hopping model of SQUID in the off-state.
The supercurrent is obtained as the integral over energy of the
spectral supercurrent.
We show on Fig.~\ref{fig:continuum} a typical variation of the
spectral supercurrent as a function of energy. We find practically
no contribution of energies larger than $|\Delta_0|$ in absolute value,
as opposed to other cases such as Ref.~[\onlinecite{Martin-cont}].
We conclude that the contribution of the continuum is negligible
for the hopping model of SQUID in the off-state in which the hopping elements
are energy-independent. The supercurrent is therefore well 
approximated by Eq.~(\ref{eq:thermo}) as in Secs.~\ref{sec:DC}
and~\ref{sec:multi}. As a physical interpretation, Andreev bound states
are localized in the superconductor in a region of size set by the coherence
length. A single channel
weak link coupling two superconductors \cite{Cuevas} is a very localized
perturbation which does not couple most of the extended states in the
superconducting electrodes, which explains why the states of the
continuum are almost insensitive to the phase difference between
the superconductors in the considered geometry with localized
interfaces.

\subsection{SQUIDs involving multichannel contacts}
\label{sec:multi}
\begin{figure}[htbp]
  \begin{center}
    \includegraphics [width=.5 \linewidth]{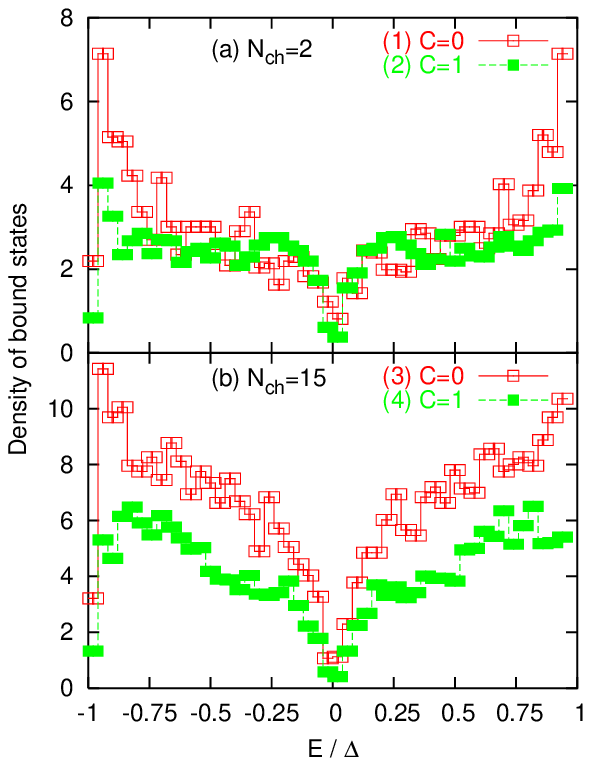}
       \caption{(Color online.) Distribution of the Andreev bound state
	 density of states
	 at phase difference $\Delta \varphi=\pi$,
	 for $N_{ch}=2$ channels
	 (a) and for $N_{ch}=15$ channels (b), without non local processes
	 (${\cal C}_0=0$, $\Box$) and with non local processes
	 to their maximal value (${\cal C}_0=1$, $\blacksquare$). 
	 Each histogram is normalized to the number of
	 Andreev bound states averaged over the realizations of the
	 Fermi phase factors.
       }
       \label{fig:dos}
  \end{center}
\end{figure}
\label{sec:multi-ins}

A metallic carbon nanotube consists of two conduction channels,
and it is thus natural to extend the discussion in Sec.~\ref{sec:DC}
to the case of multichannel\cite{Averin}
junctions (see Figs.~\ref{fig:squid dc}c and d).
We evaluate the density of Andreev bound states \cite{Lambert-ABS} for multichannel
systems with $N_{ch}=2$ and $N_{ch}=15$ channels, and with
${\cal C}_0=0$ and ${\cal C}_0=1$ (see Fig.~\ref{fig:dos}).
Increasing the strength of non local processes by increasing
${\cal C}_0$ reduces the density of Andreev bound states,
therefore reducing the value of the supercurrent,
in agreement with Sec.~\ref{subsec:Amplitude}.

\section{Conclusions}\label{sec:Conclusions}

To conclude, we have discussed signatures of non local
Andreev reflection on the current-phase relation of a dc-SQUID in
a fork geometry similar to Cleuziou {\it et al.}\cite{Cleuziou}, but
with the dimension of the middle superconductor comparable to
the superconducting coherence length, so that quasiparticles
may tunnel through the superconductor, with or without electron-hole
conversion. Compared to a geometry consisting of two parallel normal
bridges connecting two superconductors \cite{Melin-Peysson,Melin-PRB-SQUID},
we investigated here processes of higher order that are not washed out
by disorder.
For idealized single channel
systems with sharp step-function variations of the superconducting
gap in the nanotube, we found that non local processes induce sample
to sample fluctuations of the current phase relation
due to the dependence of non local processes on the Fermi phase factors.
Multichannel systems capture
moderate interface transparencies between the nanotube and the
superconductor because in this case electrons incoming in 
the superconductor can propagate in the nanotube before undergoing crossed
Andreev reflection.
Increasing the strength of 
non local processes reduces the supercurrent, as for single
channel systems.

From the point of view of future experiments,
non local processes play a role in SQUIDs with fork geometries and
with junctions made of
carbon nanotubes, normal metals or semiconducting
quantum wires. It
would be interesting to measure the reduction 
of the current-phase relation of the SQUID
upon increasing the strength of crossed processes in multichannel systems,
via a comparison of samples with different dimensions, or via the
temperature dependence of the coherence length.
A more difficult experiment consists in probing
sample to sample fluctuations of the
SQUID supercurrent. A good characterization of the sample parameters
(such as number of channels, interface transparencies) is 
required in order to distinguish
between the intrinsic fluctuations of crossed processes and
unwanted variations
of the junction parameters when changing from one sample to another.
As pointed out to us by F. Giazotto, the strength of non local processes
can be monitored by the temperature dependence of the
superconducting coherence length $\xi_0=\hbar v_F/|\Delta_0|$
(in the ballistic limit) or $\xi_0=\sqrt{\hbar D/|\Delta_0|}$
(with $D$ the diffusion coefficient in the diffusive limit)
because the superconducting gap decreases
with increasing temperature.
Increasing temperature has thus the effect of reducing $|\Delta_0|$
and 
enhancing the coherence length. It is
expected that the total
amplitude of supercurrent decreases with increasing
temperature, but the relative contribution of non local processes increases
upon increasing temperature.

\section{Acknowledgments}
The authors thank V. Bouchiat and W. Wernsdorffer for useful
discussions on their experiments, and thank
S. Florens, M. Houzet, Th. Jonckheere,
Th. Martin for useful suggestions and crucial discussions.
The authors thank D. Feinberg for useful discussions having resulted in
this work, and for stimulating discussions.

\appendix

\section{Fluctuations of non local
transport through a diffusive superconductor}
\label{app:osc-diff}
The charge transmission coefficient of a diffusive superconductor
vanishes after averaging over disorder
in the diffusive limit 
\cite{Feinberg-des,Duhot-Melin}.
To show that the charge transmission coefficient of a disordered
superconductor
fluctuates at the scale of the Fermi wave-length $\lambda_F$, we
show that the opposite hypothesis does not hold.
Simply, we obtain fluctuations of the charge transmission coefficient
by adding a very small extra ballistic region 
much larger than the Fermi wave-length but much smaller than the
elastic mean free path.

\section{Critical current in the tunnel limit}
\label{app:pert}
In this Appendix,
the supercurrent is expanded in the tunnel amplitude $t_{a,b}$
connecting the two interfaces between the superconductors.
We do not detail the corresponding calculation based on 
diagrammatic perturbation theory.
We start with
the first terms of an expansion of the supercurrent
in the tunnel amplitudes [see the
following Eqs.~(\ref{eq:IS})-(\ref{eq:toto1})].
The dimensionless parameters $\tau_{a,b}=(\pi\rho_N t_{a,b})^2$,
related in the tunnel limit to the dimensionless interface
transparencies through $T_{a,b}\simeq4\tau_{a,b}$,
are supposed to be small. The Dyson equations are expanded systematically
in the tunnel amplitudes and the lowest order diagrams are collected,
leading to the following expansion for the supercurrent:
\begin{eqnarray}
\label{eq:IS}
I_S(\Delta \varphi,\phi)&=&A_a \sin{(\Delta\varphi+\phi)}
+A_b\sin{(\Delta\varphi-\phi)}
+B\sin{(2\Delta\varphi)}\\
&+&D_a\sin{(2(\Delta\varphi+\phi))}
+D_b\sin{(2(\Delta\varphi-\phi))}
\nonumber
,
\end{eqnarray}
with
\begin{eqnarray}
\label{eq:Aa}
A_a&=&\frac{e}{h}\tau_a|\Delta_0|\\
\label{eq:Ab}
A_b&=&\frac{e}{h} \tau_b|\Delta_0|\\
\label{eq:sin2}
B&=&-\frac{e}{2h}
\tau_a\tau_b {\cal C}_*^2 |\Delta_0| \sin^2{(k_F R_{\alpha,\beta})}\\
D_a&=&-\frac{e}{2h}\tau_a^2|\Delta_0|\\
D_b&=&-\frac{e}{2h}\tau_b^2|\Delta_0|
,
\end{eqnarray}
where ${\cal C}_*$ is the value of the parameter ${\cal C}(\omega)$ 
where $\hbar \omega$ takes the value of the bound state energy level.
In this case, the bound states are very close to the gap, so that
${\cal C}_*\simeq 1$.
The contribution of non local Andreev reflection to the supercurrent
(term $B$ in Eq.~(\ref{eq:sin2})) is, in the tunnel limit,
much smaller than the contribution of local tunneling of Cooper
pairs from one superconductor to the other
(terms $A_a$ and $A_b$ in Eqs.~(\ref{eq:Aa}) and (\ref{eq:Ab})).

The reduction of the supercurrent upon including non local processes
is described in the tunnel limit by an expansion of Eq.~(\ref{eq:IS})
around $\phi=\pi/2$: $\phi=\pi/2+\delta\phi$. 
We consider an ensemble of
single channel systems with a collection of $k_F R_{\alpha,\beta}$
and thus we average to $1/2$ the
factor $\sin^2{(k_F R_{\alpha,\beta})}$
in Eq.~(\ref{eq:sin2}).
We define $\Delta\varphi^{(0)}$ as the value of $\Delta \varphi$
such that
\begin{equation}
\frac{\partial I_S(\Delta\varphi=\Delta \varphi^{(0)},\phi)}
{\partial(\Delta \varphi)}=0
,
\end{equation}
and we obtain
\begin{equation}
\cos{(\Delta \varphi^{(0)})}=
\frac{(1-\tau)\delta\phi\pm\sqrt{
((1-\tau)\delta\phi)^2+2\tau^2(2-{\cal C}_*^2/2)^2}}
{2\tau(2-{\cal C}_*^2/2)^2}
,
\end{equation}
where we supposed a symmetric contact with $\tau_{a,b}=\tau$.
For $\delta\phi\gg\tau$, the critical current is given by
\begin{equation}
I_c(\pi/2+\delta\phi)=4\pi\tau|\Delta_0|\delta\phi
,
\end{equation}
and for $\delta\phi\ll\tau$, it is given by
\begin{equation}
\label{eq:toto1}
I_c(\pi/2+\delta\phi)=\pi\tau^2|\Delta_0|
\left(2-\frac{{\cal C}_*^2}{2}\right)
.
\end{equation}
For $\phi=0$ the expansion of the critical current is given by
\begin{equation}
\label{eq:TOTO}
I_c(0)= \frac{2\Delta e\tau}{h}+ {\cal O} (\tau^3),
\end{equation}
where non local effects do not enter Eq.~(\ref{eq:TOTO}) at the leading order.
We conclude in the tunnel limit to a reduction
of the contrast of the
critical current oscillations upon increasing the strength ${\cal C}_*$ of
non local processes.
The main body of the article corresponds to interfaces
with moderate of large transparencies, described by Andreev bound
states obtained from the Green's function dressed by tunnel processes
to infinite order, as opposed to the limit of tunnel contacts considered
in this Appendix.

\section{Andreev bound states}\label{sec:Appendice B}
\label{app:abs}
In a symmetric SQUID with $t_a=t_b$ and no
magnetic field ($\phi=0$), the bound states take the form 
\begin{eqnarray}\label{omega sym pm}
\Omega^{\pm}_{1}(\Delta\varphi)=\pm
|\Delta_0| \sqrt{A^+(\Delta\varphi)}/\sqrt{B^+(\Delta\varphi)}
\end{eqnarray}
\begin{eqnarray}\label{omega antisym pm}
\Omega^{\pm}_{2}(\Delta\varphi)=\pm
|\Delta_0| \sqrt{A^-(\Delta\varphi)}/\sqrt{B^-(\Delta\varphi)}
\end{eqnarray}
 with 
\begin{eqnarray}\label{A}
  A^{\pm}(\Delta\varphi)&=&2\cos(\Delta\varphi)T\left(1\pm
  {\cal C}_*\sin(k_FR_{\alpha,\beta})\right)+1\\
  &&+T^2\left(1\pm {\cal C}_*
  \sin(k_FR_{\alpha,\beta})\right)^2 +T^2{\cal
  C}_*^2\cos(k_FR_{\alpha,\beta})^2
\nonumber
\end{eqnarray}
and
\begin{eqnarray}\label{B}
  B^{\pm}(\Delta\varphi)&=&\left[1+T\left(1\pm
  {\cal C}_*\sin(k_FR_{\alpha,\beta})\right)\right]^2+T^2{\cal
  C}_*^2\cos(k_FR_{\alpha,\beta})^2
\end{eqnarray}
where $\tau=\pi^2\rho_N^2t^2=t^2/W^2$,
related to the normal transmission by the 
relation \cite{Cuevas}
\begin{eqnarray}\label{normal transmission}
T_{NN}^{a,b}=(4t^2_{a,b}/W^2)/(1+t^2_{a,b}/W^2)^2
\end{eqnarray}
where $W=1/\pi\rho_N$ is the band-width,
with $\rho_N$ the normal density of states. 
The bound state levels are determined self-consistently
in such a way as ${\cal C}_*$ is the value of
${\cal C}(\omega)$ [see Eq.~(\ref{eq:C-omega})],
where $\hbar \omega$ is replaced by the bound state energy
in a self-consistent manner.

Lets us consider the case $k_FR_{\alpha,\beta}=
2\pi n$ (with $n$ an integer). The bound states
levels deduced from Eqs (\ref{A}) and (\ref{B}) are then degenerate:
\begin{eqnarray}\label{Omega sym}
\Omega^-_{1}(\Delta\varphi)=\Omega^-_{2}(\Delta\varphi)
=-|\Delta_0|\sqrt{1-\alpha
\sin(\Delta\varphi/2)^2}
\end{eqnarray}
with
\begin{eqnarray}
\alpha&=&4\tau/((1+\tau)^2+\tau^2{\cal C}_*^2)
=\frac{4t^2/W^2}{(1+t^2/W^2)^2+{\cal C}_*^2t^4/W^4}.
,
\end{eqnarray}
where we introduced the band-width $W$ according to Ref.~[\onlinecite{Cuevas}].
The SQUID is then equivalent to two identical S-I-S junctions, as seen from
comparing Eq. (\ref{Omega sym}) to Ref.~[\onlinecite{Cuevas}]. 

For $k_FR_{\alpha,\beta}=\pi/2+2\pi n$ (with $n$ an integer) the degeneracy is
removed only if ${\cal C}_0\neq 0$:
\begin{eqnarray} 
\Omega_{1,2}^-(\Delta\varphi)
=-|\Delta_0|\sqrt{1-\alpha^{(1,2)}\sin(\Delta\varphi/2)^2},
\end{eqnarray}
with
\begin{eqnarray}
\alpha^{(1,2)}=\frac{4\tau(1\pm{\cal C}_*)}{(1+\tau(1\pm{\cal C}_*))^2}
,\end{eqnarray}
where the ``$+$''  and ``$-$'' signs correspond to ``1'' and ``2''
respectively.

In the presence of a magnetic field ($\phi\neq 0$), the bound state
energy levels take the form
\begin{eqnarray}
\Omega_{1,2}^-(\Delta\varphi)=
 -\tilde{|\Delta_0|}\sqrt{1-\alpha^{(1,2)}\sin(\Delta\varphi/2)^2-\beta^{(1,2)}
\sin(\Delta\varphi)}
\end{eqnarray}
with 
\begin{eqnarray}
\alpha^{(1,2)}=
\frac{4\tau\cos(2\phi)(1\pm{\cal C}_*)}{(1+\tau(1\pm{\cal
    C}))^2-4\tau\sin(\phi)^2(1\mp{\cal C}_*)\pm 4\tau^2
{\cal C}_*\sin(2\phi)^2},
\end{eqnarray}
\begin{eqnarray}
\beta^{(1,2)}=
\frac{2\tau\sin(2\phi)(1\mp{\cal C}_*)}{(1+\tau(1\pm{\cal
    C}))^2-4\tau\sin(\phi)^2(1\mp{\cal C}_*)\pm 4\tau^2{\cal C}_*
\sin(2\phi)^2}.
\end{eqnarray}
and
\begin{eqnarray}
\tilde{|\Delta_0|}=|\Delta_0|\sqrt{1-4\tau\frac{\sin(\phi)^2(1\mp{\cal
    C})\pm \tau{\cal C}_*\sin(2\phi)^2}{(1+\tau(1\pm{\cal C}_*))^2}}
.
\end{eqnarray}



\begin{thebibliography}{99}
\bibitem{Beckmann} D. Beckmann, H.B. Weber and H.v. L\"ohneysen,
 Phys. Rev. Lett. {\bf 93}, 197003 (2004).
  
\bibitem{Russo} S. Russo, M. Kroug, T.M. Klapwijk and A.F. Morpurgo,
 Phys. Rev. Lett. {\bf 95},  027002 (2005).

\bibitem{Cadden} P. Cadden-Zimansky and V. Chandrasekhar, cond-mat/0609749.

\bibitem{Byers} J. M. Byers and M. E. Flatt\'e,  Phys. Rev. Lett.
  {\bf 74}, 306 (1995).
  
\bibitem{Deutscher} G. Deutscher and D. Feinberg,
  App. Phys. Lett. {\bf 76},
  487 (2000);
  
\bibitem{Falci}  G. Falci, D. Feinberg and F. Hekking,
  Europhysics Letters {\bf 54}, 255 (2001).

\bibitem{Samuelson} P. Samuelsson, E.V. Sukhorukov and M. B\"uttiker,
Phys. Rev. Lett. {\bf 91}, 157002 (2003).

\bibitem{Prada} E. Prada and F. Sols, Eur. Phys. J. B
\textbf{40}, 379 (2004).

\bibitem{Koltai} P.K. Polin\'{a}k, C.J. Lambert, J. Koltai and
J. Cserti, Phys. Rev. B {\bf 74}, 132508 (2006).

\bibitem{japs} T. Yamashita, S. Takahashi and S. Maekawa,
Phys. Rev. B {\bf 68}, 174504 (2003).

\bibitem{Feinberg-des} D. Feinberg, Eur. Phys. J. B
  {\bf 36}, 419 (2003).

\bibitem{Melin-Feinberg-PRB} R. M\'elin and D. Feinberg,
  Phys. Rev. B {\bf 70}, 174509 (2004)

\bibitem{Melin-PRB} R. M\'elin, 
  Phys. Rev. B  {\bf 73}, 174512 (2006).

\bibitem{Levy} A. Levy Yeyati, 
F.S. Bergeret, A. Martin-Rodero and T.M. Klapwijk,
cond-mat/0612027.

\bibitem{Duhot-Melin} S. Duhot and R. M\'elin, Eur. Phys. J. B
{\bf 53}, 257 (2006).

\bibitem{Morten} J.P. Morten, A. Brataas and W. Belzig,
Phys. Rev. B {\bf 74}, 214510 (2006).

\bibitem{Giazotto} F. Giazotto, F. Taddei, F. Beltram and
R. Fazio, Phys. Rev. Lett. {\bf 97}, 087001 (2006).

\bibitem{Golubov} A. Brinkman and A.A. Golubov, 
 Phys. Rev. B {\bf 74}, 214512 (2007).

\bibitem{Zaikin} M.S. Kalenkov and A.D. Zaikin, cond-mat/0611330.

\bibitem{Duhot-Melin-cond-mat} S. Duhot and R. M\'elin,
cond-mat/0710748.

\bibitem{Choi} M.S. Choi, C. Bruder and D. Loss, Phys. Rev. B {\bf 62},
  13569 (2000); P. Recher,E. V. Sukhorukov and D. Loss,
Phys. Rev. B {\bf 63}, 165314 (2001).

\bibitem{Martin} G. B. Lesovik, T. Martin and G. Blatter,
Eur. Phys. J. B \textbf{24},
287 (2001); N. M. Chtchelkatchev,
G. Blatter, G.B. Lesovik and T. Martin,
Phys. Rev. B \textbf{66}, 161320(R) (2002).

\bibitem{Bouchiat} V. Bouchiat, N. Chtchelkatchev, D. Feinberg,
G.B. Lesovik, T. Martin and J. Torres,
Nanotechnology {\bf 14}, 77 (2003).

\bibitem{Melin-Peysson} R. M\'elin and S. Peysson,
Phys. Rev. B {\bf 68}, 174515 (2003) 

\bibitem{Melin-PRB-SQUID} R. M\'elin, 
Phys. Rev. B {\bf 72}, 134508 (2005).

\bibitem{Delft} P. Jarillo-Herrero, J.A. van Dam and L.P. Kouwenhoven,
Nature {\bf 439}, 953 (2006).

\bibitem{Cleuziou} J. P. Cleuziou, W. Wernsdorfer, V. Bouchiat,
  T. Ondarçuhu, M. Monthioux, Nature Nanotechnology {\bf 1}, 53 (2006).

\bibitem{note-xi} The coherence length at zero energy
is $\xi_{0,ball}=\hbar v_F/|\Delta_0|$
in a ballistic system (with $v_F$ the Fermi velocity and $|\Delta_0|$
the superconducting gap). In a diffusive system,
$\xi_{0,diff}\sim\sqrt{\xi_{0,ball}l_e}$, with $l_e$ the elastic
mean free path.

\bibitem{Cleuziou2} J.-P. Cleuziou, W. Wernsdorfer, V. Bouchiat,
  Th. Ondarcuhu, M. Monthioux, cond-mat/0610622 (2006)

\bibitem{Andreev} A.F. Andreev, Zh. Eksp. Teor. Fiz. {\bf 46},
1823 (1964) [Sov. Phys. JETP {\bf 19}, 1228 (1964)].

\bibitem{Caroli} C. Caroli, R. Combescot, P. Nozi\`eres, D. and
  Saint-James, J. Phys. C: Solid St. Phys. {\bf 4}, 916 (1971); {\it
  ibid} {\bf 5}, 21, (1972).

\bibitem{Cuevas} J. C. Cuevas, A. Mart\'in-Rodero, and A. Levy Yeyati,
  Phys. Rev. B {\bf 54}, 7366 (1996);
J. C. Cuevas, A. Martín-Rodero, and A. Levy Yeyati
Phys. Rev. Lett. 82, 4086 (1999).

\bibitem{Averin} D. Averin and A. Bardas
Phys. Rev. Lett. {\bf 75}, 1831 (1995);
D. Averin and H. T. Imam
Phys. Rev. Lett. 76, 3814 (1996).

\bibitem{Hekking} F.W.J. Hekking and Yu. V. Nazarov, Phys. Rev. Lett.
{\bf 71}, 1625 (1993); Phys. Rev. B {\bf 44}, 11506 (1991).

\bibitem{Lambert-ABS} A. Kormanyos, Z. Kaufmann, J. Cserti, C.J. Lambert,
Phys. Rev. Lett. {\bf 96}, 237002 (2006) and references theresin.

\bibitem{Beenakker} C.W.J. Beenakker, Phys. Rev. Lett. {\bf 67}, 3836 (1991).

\bibitem{Bardeen} J. Bardeen, R. K\"ummel, A.E. Jacobs and L. Tewordt,
Phys. Rev. {\bf 187}, 556 (1969).

\bibitem{AGD} A.A. Abrikosov, L.P. Gorkov and I.E. Dzyaloshinski,
{\it Methods of Quantum Field Theory in Statistical Physics}
(Dover, New York, 1963).

\bibitem{Martin-cont} C. Benjamin, Th. Jockqueere,
A. Zazunov and Th. Martin, cond-mat/0605338.

\end{thebibliography}
\end{document}